\documentclass[11pt]{article}
\usepackage{moriond,epsfig,amssymb,amsmath,multirow}
\usepackage{graphicx}
\usepackage{float}

\newcommand{\bea}{\begin{eqnarray}}
\newcommand{\eea}{\end{eqnarray}}
\newcommand{\ie}{{\it ie}}
\newcommand{\xds}{extra-dimensions }
\newcommand{\Xds}{Extra-dimensions }
\newcommand{\xd}{extra-dimension }

\newcommand{\djs}{di-jets }
\newcommand{\Djs}{Di-jets }
\newcommand{\djj}{di-jet }

\newcommand{\as}{$\alpha_S$ }
\newcommand{\xdpar}{$(\delta,M_c)$ }
\renewcommand{\to}{\rightarrow}

\input{epsf}

\begin{document}
\vspace*{4cm}

\title    {  PROTON STRUCTURE IMPACT ON SENSITIVITY TO EXTRA-DIMENIONS 
       AT LHC }

\author{ S. FERRAG }
\address{ATLAS Collaboration\\ Laboratoire de Physique Nucl\'eaire et des hautes Energies,
Paris, France}
\maketitle\abstracts{   
 The LHC data will provide sensitivity to an unification of the couplings at low energies 
in the range $\sim$10-100 TeV. It is demonstrated in this note that the lack of knowledge on the proton 
structure, specifically its gluon distribution, can lower dramatically the sensitivity of bare cross 
section measurements to this physics. However, some more elaborated strategies could probably be developped 
to recover an important part of the sensitivity.}

  

\section{Introduction}

Theoretical particle physics deals with three fundamental energy scales: the electroweak breaking
scale at $\approx 10^2$ GeV, the grand unification theory (GUT) scale at  $\approx 10^{16}$
GeV and the Planck scale, where gravitation becomes as strong as the other interactions,
 at $\approx 10^{19}$ GeV. Those 
three scales are separated by a troubling physical energy range in which the Standard Model  (SM) predicts no new physics. 
This forms the well known hierarchy problem. A new framework has been recently proposed to adress this problem 
\cite{ArkaniHamed:1998rs,Randall:1999ee}.
Related models contain only one fundamental scale, which can be as low as few TeV, and a more complex space-time structure
with $\delta$ new compactified spacial dimensions.
In this picture, the Standard Model matter fields are localised in a (3+1) dimensional hyperplane (3-brane). The above scales
become effective and can be expressed as functions of the fundamental scale and the compactification radius 
of the \xds \cite{ArkaniHamed:1998rs}. This new picture induces 
interesting phenomenological aspects at the LHC if the fundamental scale is as low as a few TeV (10-100 TeV): 
production of gravitons 
and observation of Kaluza Klein (KK) excitations will be possible if the radius size of one of the $\delta$ 
compactified \xds is
about a few TeV. If gauge bosons can propagate in the \xds, we also expect a violation of the SM logarithmic 
behaviour of the running couplings \cite{Dienes:1999vg,Dienes:1998vh}.
This note will concentrate on this latter aspect. LHC data will provide a measurement
of the different couplings on a large energy range and allow for experimental sensitivity to a possible non SM running. 
Following reference \cite{balazs}, the measurement of \as is investigated by looking at the
\djs cross section.\\

In this analysis, the \djs cross section in the presence of \xds is studied and QCD aspects of this measurement are investigated. \Xds affect
this cross section through the \as evolution. This new \as running is given in references 
\cite{Dienes:1998vh,Dienes:1999vg}. Kaluza Klein excitations are included only in the \as running through 
the $\beta$ coefficients in the Renormalization Group Equations (RGE). 
The explicit exchange of these excitations is not taken into account in the calculation of the \djs cross
section.

A preliminary work on the sensitivity to \xds\noindent at the LHC, using \djs production 
\cite{balazs}, shows that the sensitivity to \xds at LHC reaches 5 to 10 TeV in compactification scale 
following the \as running scenario (see below).\\

A complementary analysis using the Monte Carlo techniques is presented here:
a first step consisted in the implementation of the new \as running in the event generator Pythia 6.152. 
As the \djs production
will be done over a large Pt range, the impact of Parton Density Function (PDF) 
uncertainties in the proton on the sensitivity of the di-jets cross 
section to Standard Model RGE violations is investigated. 
This is done using PDFs from CTEQ6 that come with 40 PDF sets. This framework
enables, for the first time, to investigate rigourously the impact of the PDF 
uncertainties on a given measurement. 
The results are then expressed in terms of an  
impact of the proton structure on the \xds sensitivity.  

\begin{figure}[t]
\unitlength 1mm 
\begin{center}
\leavevmode
\epsfig{file=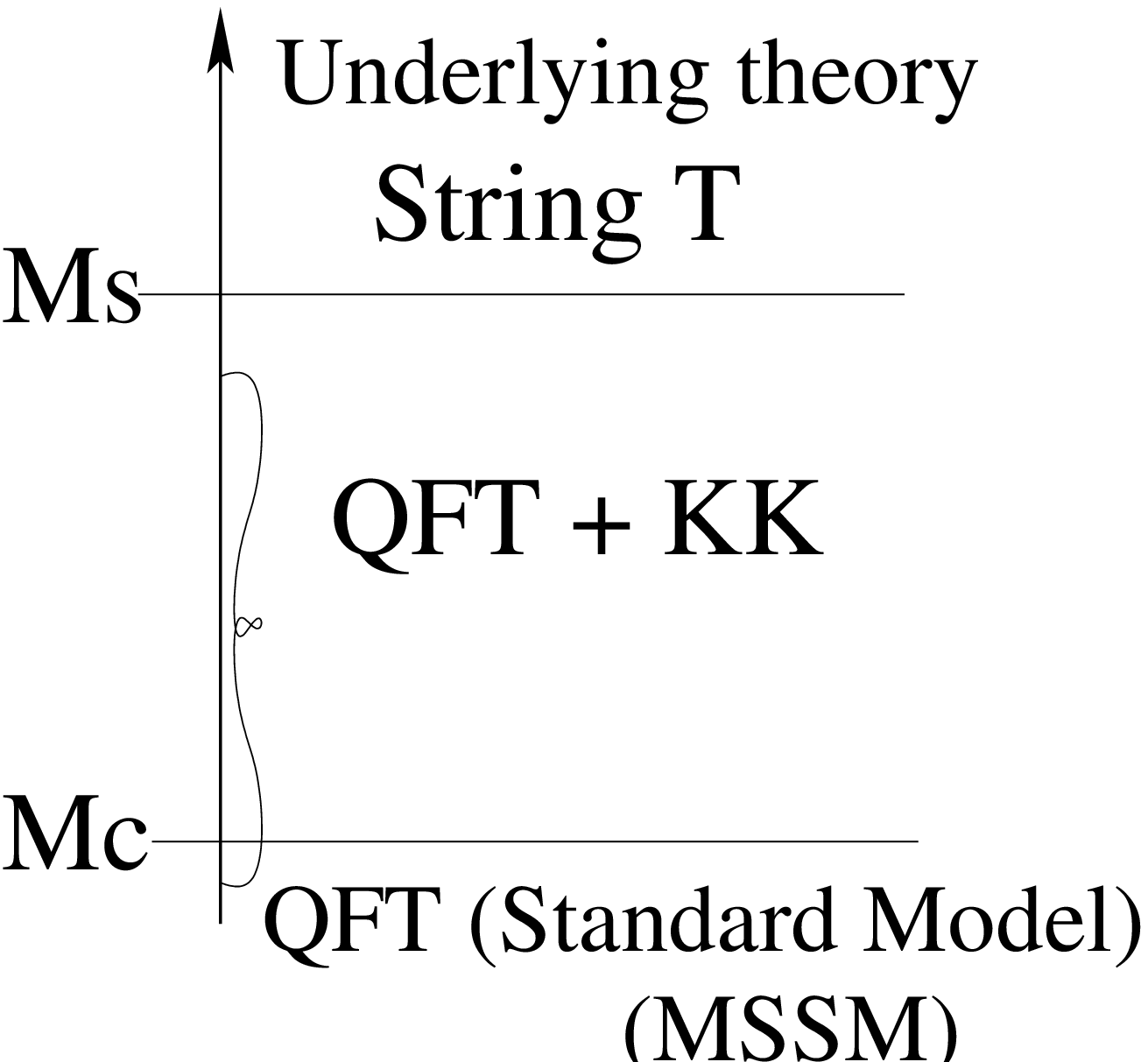,width=0.5\linewidth}
\caption{\it\label{fig:qft_string} In the transition region between the underlying theory scale around the
string scale $M_s$ and the Standard Model scale below the compactification scale $M_c$, the underlying
theory is approached with a Quantum Field Theory (QFT) including the KK excitation. The validity region
of calculation given in references [3,4] extends up to this transition region.}
\end{center}
\end{figure}


\section{Event Generation}
 
The \djs cross section in the \xds regime is assumed to be 
a continuity of the Standard Model \djs cross section with a new \as  running. This can be written as:
\bea
\frac{d\sigma^{XD}}{d M_{JJ} }=\frac{d\sigma^{SM}}{d M_{JJ}}(\alpha_s^{XD}),
\eea 
where $\sigma^{XD}$ and $\alpha_s^{XD}$ are respectively, the cross section and the strong coupling 
in presence of \xds and $\sigma^{SM}$ is the cross section computed within the Standard Model. $M_{JJ}$ is 
the di-jets invariant mass.\\

The implemented model is inspired from references \cite{Dienes:1998vh,Dienes:1999vg}. We
assume that the underlying theory has 3+$\delta$ independent space dimensions and one time dimension.
The $\delta$ \xds are compactified on a hypersphere of radius $R_c$. In this theory all the 
fundamental scales are close to the weak scale and standard gauge bosons propagate in the \xds.
The compactification scale $M_c$ is between two scales
\bea
M_{Z^0} \lesssim M_c = 1/R_c \lesssim M_{string} = {\cal{O}}(100 \mbox{ \rm TeV}). 
\eea
Following \cite{Dienes:1998vh,Dienes:1999vg}, we assume that in the TeV energy range  
the theory can be well approximated by a field 
theory formulated in a 4-dimensional space-time ( figure \ref{fig:qft_string}). In these papers, the field theory 
is assumed to be the Minimal Supersymmetric Standard 
Model (MSSM) whose RGE have no sensitive
 difference with the Standard Model ones at few TeV scale\footnote{The difference of the gauge evolutions
between the Standard Model and the MSSM is negligible compared to the effect of
an O(TeV) size extra dimension in the $\mu$ = 1-10 TeV range.  Keeping in agreement
with  References \cite{Dienes:1998vh,Dienes:1999vg}, the MSSM beta functions are used.}.  
In this study, the field theory is supposed to be
the Standard Model. Consequently, 
no supersymmetric particles
are put in the physical spectrum, only \as MSSM running is used.
The presence of the KK excitations of the gluons, photons, $Z^0$ and $W^\pm$ affects the renormalization 
evolution of the gauge coupling by 
power-law type corrections. At the lowest order, the scale dependence of the gauge couplings is given by\cite{Dienes:1998vh,Dienes:1999vg}:
\begin{eqnarray}
       \alpha_i^{-1}(\mu) = \alpha_i^{-1}(\mu_0) -
            {b_i-\tilde b_i\over 2\pi}\,\ln{\mu \over \mu_0} 
          -~{\tilde b_i\over 4\pi}\,
             \int_{r\mu^{-2}}^{r\mu_0^{-2}} {dt\over t} \,
     \left[ \vartheta_3\left( {it\over \pi R_c^2} \right) \right]^\delta,
\label{KKresult}
\end{eqnarray}
where $i = 1,2,3$ labels the gauge groups of the MSSM, and the coefficients of
the usual one loop beta functions:
\begin{eqnarray}
        (b_1,b_2,b_3) = (33/5,1,-3),
\end{eqnarray}
are supplemented by new contributions from the properly supersymmetrized 
KK excitations
\begin{eqnarray}
      (\tilde b_1,\tilde b_2,\tilde b_3) = (3/5, -3, -6) + \eta \; (4, 4, 4),
\end{eqnarray}
where $\eta$ is the number of chiral fermions in the theory and set to be $\eta = 0$ for simplicity. In the last term of Eq.
(\ref{KKresult}) $\vartheta_3$ denotes the elliptic Jacobi function and 
\begin{eqnarray}
      r = \pi \,(X_\delta)^{-2/\delta} ~~~ {\rm with} ~~~ 
      X_\delta = {2 \pi^{\delta/2} \over \delta \Gamma(\delta/2)}.
\end{eqnarray}

The power-law term in Eq. (\ref{KKresult}) accelerates the running of the gauge 
couplings and makes them meet earlier than the usual unification scale of
$\approx 10^{16}$ GeV. In particular, for $\eta = 0$, the strong coupling decreases 
faster than what the logarithmic running describes.

\subsection{Implementation of $\alpha_S$}

\begin{figure}[t]
\unitlength 1mm 
\begin{center}
\leavevmode
\epsfig{file=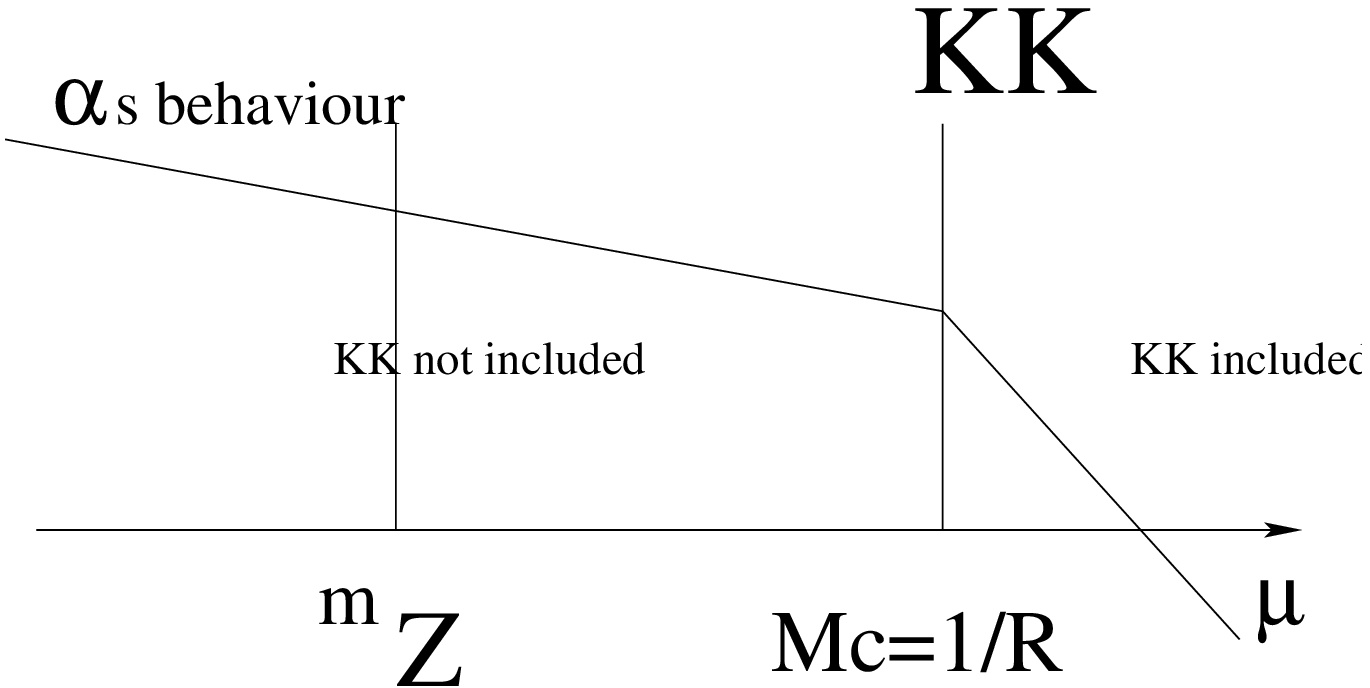,width=0.45\linewidth}
\epsfig{file=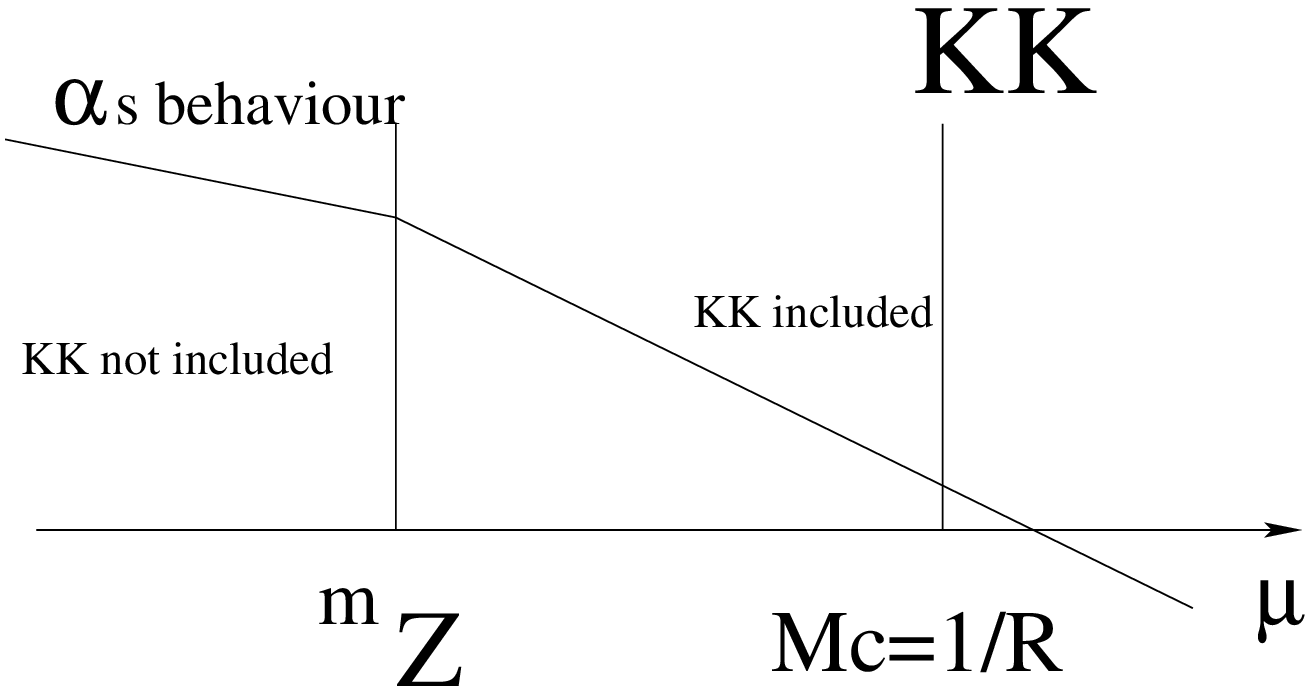,width=0.45\linewidth}
\caption{\it\label{fig:opt_pess}(Left) pessimistic scenario: the macthing between the Standard Model running of \as
and the new running is performed at the KK mass. This doesn't take into account the KK width. (Right) 
optimistic scenario: the matching between the two \as runnings is performed at the $Z^0$ mass. The KK width is 
overestimated. The matching at the KK threshold is discussed in details in reference[6]. }
\end{center}
\end{figure}

Following \cite{balazs}, the transition between the usual Standard Model running and the modified running 
can be done whithin two extreme scenarii, the so called pessimistic and optimistic ones 
(figure \ref{fig:opt_pess}). 
The pessimistic case assumes
that the running is altered only above $\mu=1/R_c$. The optimistic scenario assumes that the asymptotic
high energy region formula can be downscaled to the $Z^0$ mass in Eq. (\ref{KKresult}). 
A correct treatment of the transition should take into account Kaluza Klein width effects which are large as
$\Gamma_n = 2 \alpha_S(Q) n/R_c$\footnote{For exemple: the width of a 10 TeV resonnance is about 2 TeV.}. 
Some theoretical work is ongoing to treat this problem properly.
In this note, only the pessimistic scenario
is implemented in Pythia 6.152 and used in this analysis in a conservative way.\\

Preliminary results on the sensitivity to \xds given in \cite{balazs} show that the 
sensitivity to \xds reaches 5 TeV (10 TeV) in compactification scale using the
pessimistic (optimistic) prescription. Moreover, these results depend very smoothly on the number of 
\xds\footnote{For $Mc=5$ TeV, the separation between the Standard Model prediction and the \xds prediction is in order of 
5$\sigma$s independently from the number of \xds.}. \\ 

\begin{figure}[t]
\unitlength 1mm 
\begin{center}
\leavevmode
\epsfig{file=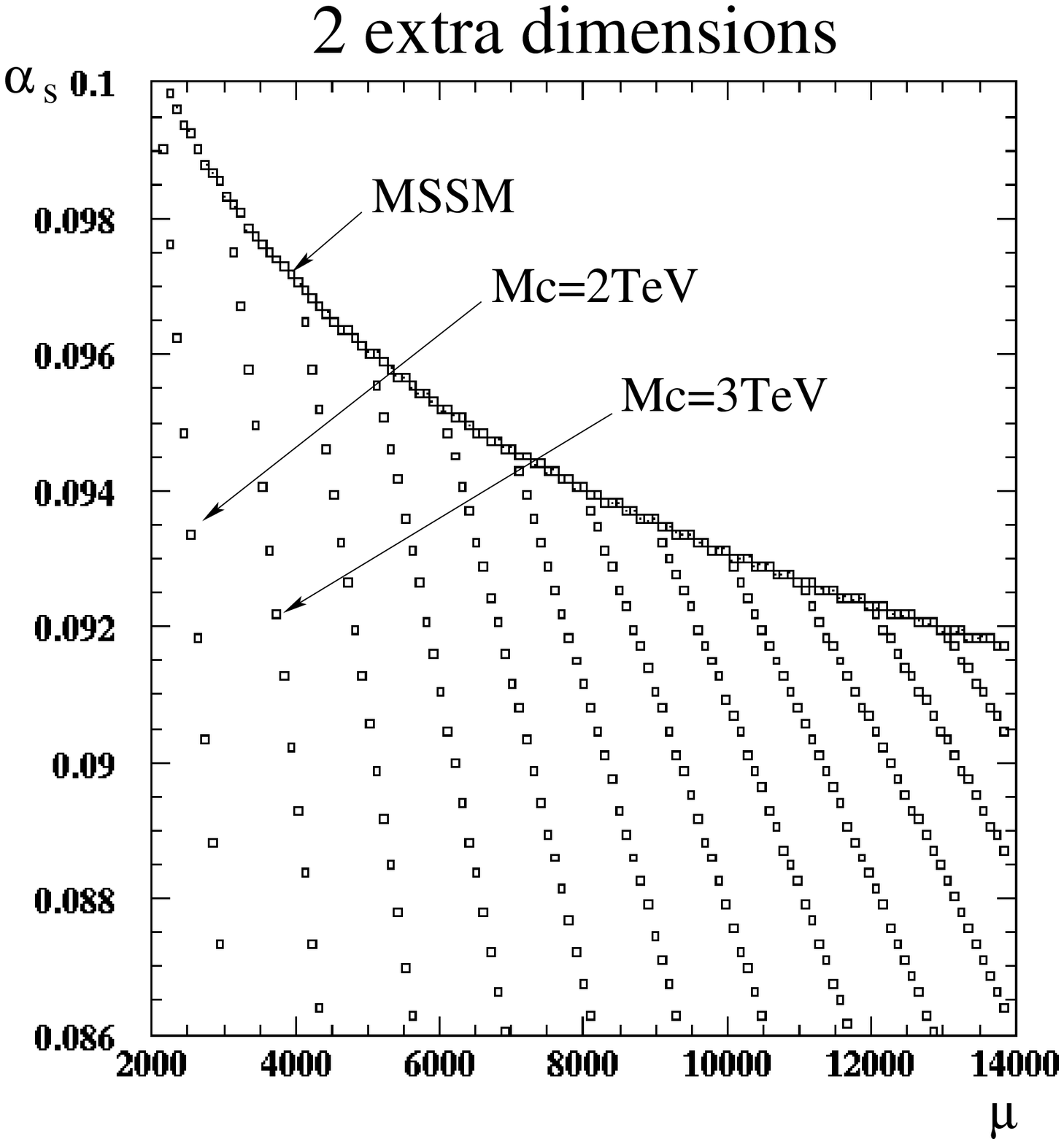,width=0.49\linewidth}
\epsfig{file=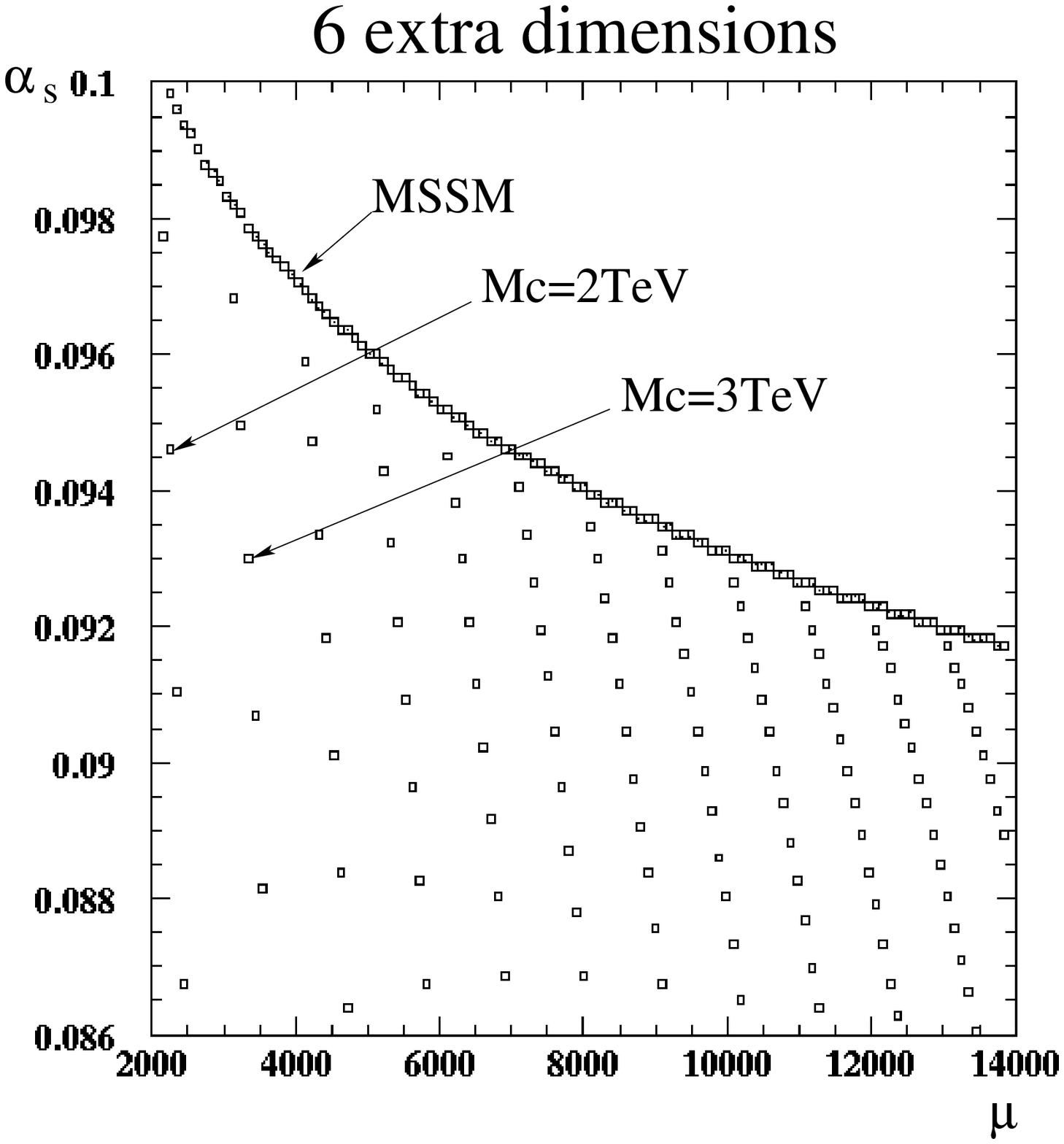,width=0.49\linewidth}
\caption{\it\label{fig:alpha_s_nxd}Implemented \as running for $\delta =2$ (left) and $\delta=6$ \xds (right). 
The upper line corresponds to the MSSM logarithmic running. Every going-down line corresponds to new running 
for a given compactification scale. This latter vary from $M_c$=2 to 14 TeV. For a given value of $M_c$,
the considered \as running is composed by the MSSM part below $M_c$ and the new part above $M_c$ with 
the pessimistic scenario matching at $M_c$.}
\end{center}
\end{figure}

Figure (\ref{fig:alpha_s_nxd}) shows the implemented running for different \xds parameters \xdpar. 
The values of the implemented compactification scale $M_c$ vary from 2 to 14 TeV 
and 2, 4 and 6 for the numbers of \xds $\delta$.
Below a given value of $M_c$, we can see the MSSM logarithmic running of \as and the \xds corrected one
above $M_c$. Figure (\ref{fig:alpha_s_nxd}.a) shows for $\delta=2$ \xds and for different 
compactification scales the \as running among the energy scale. figure (\ref{fig:alpha_s_nxd}.b) 
shows \as running for $\delta=6$ \xds\noindent.

\subsection{\Djs cross section in presence of \xds}

Once the \xds \as running implemented, di-jet events are simulated within the
ATLAS fast simulation framework. Because of the strong decreasing behaviour of the 
the cross section, events are generated in the Pt range 500-5000 GeV by step 
of $\Delta$Pt=500 GeV. The lower Pt limit $Pt_{min}$ is varied as: 
\bea
Pt_{min}=500 ~.~ i \mbox{ \rm GeV }, ~~~~~i=1 \mbox{ \rm to } 10.
\eea 
10 000 events are generated for each \xds parameter \xdpar  and for each Pt bin.\\

\begin{figure}[t]
\unitlength 1mm 
\begin{center}
\leavevmode
\epsfig{file=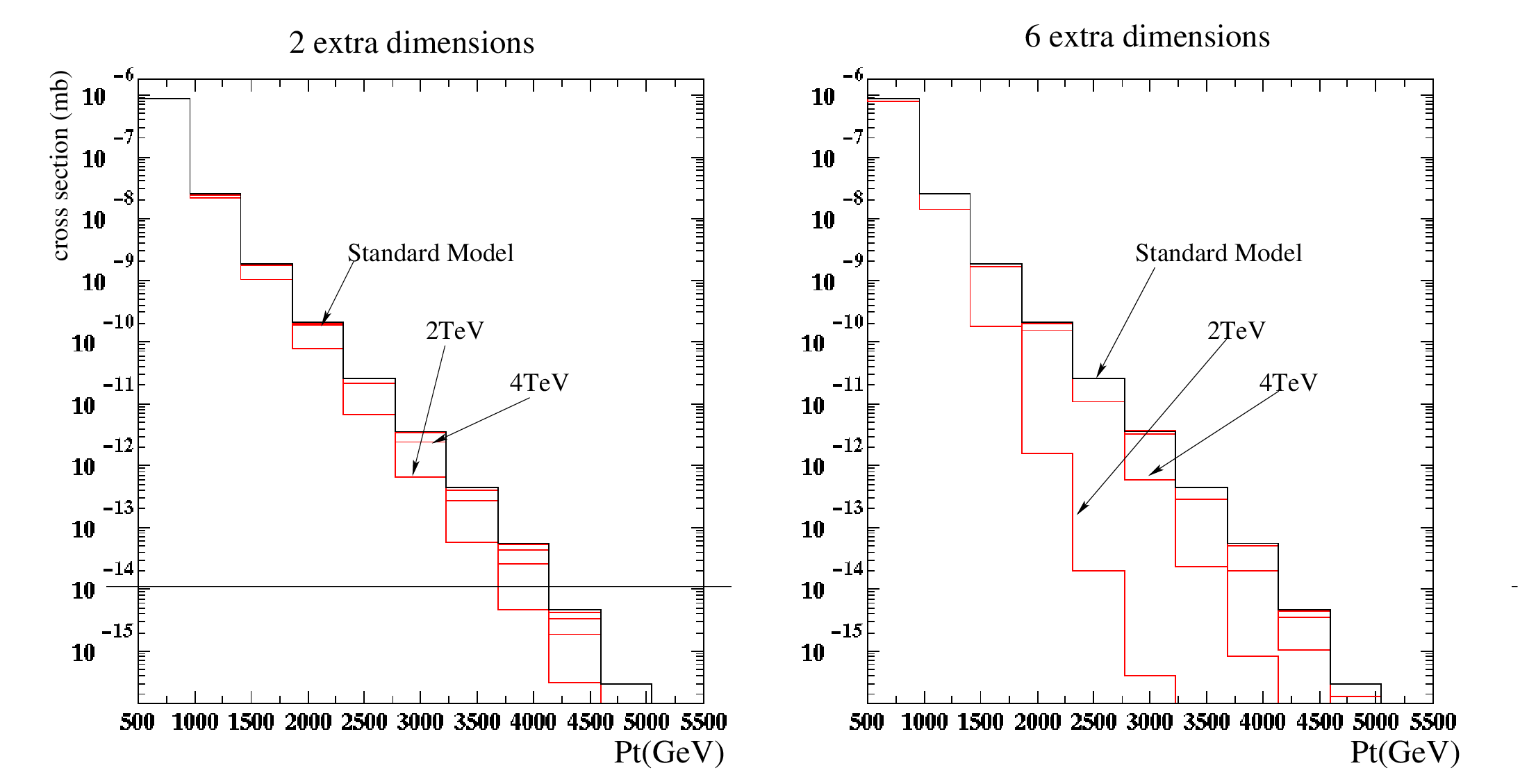,width=0.99\linewidth}
\caption{\it\label{fig:2_6xd} \djs cross section predictions for $\delta=2$ ($\delta$=6) \xds at left (at right). 
The horizontal line shows the limit corresponding to the first
year of LHC high luminosity, \ie, 100 fb$^{-1}$. 
All the \xds predictions are below the Standard Model one. The separation between the predictions increases with
the number of \xds. }
\end{center}
\end{figure}

\begin{figure}[t]
\unitlength 1mm 
\begin{center}
\leavevmode
\epsfig{file=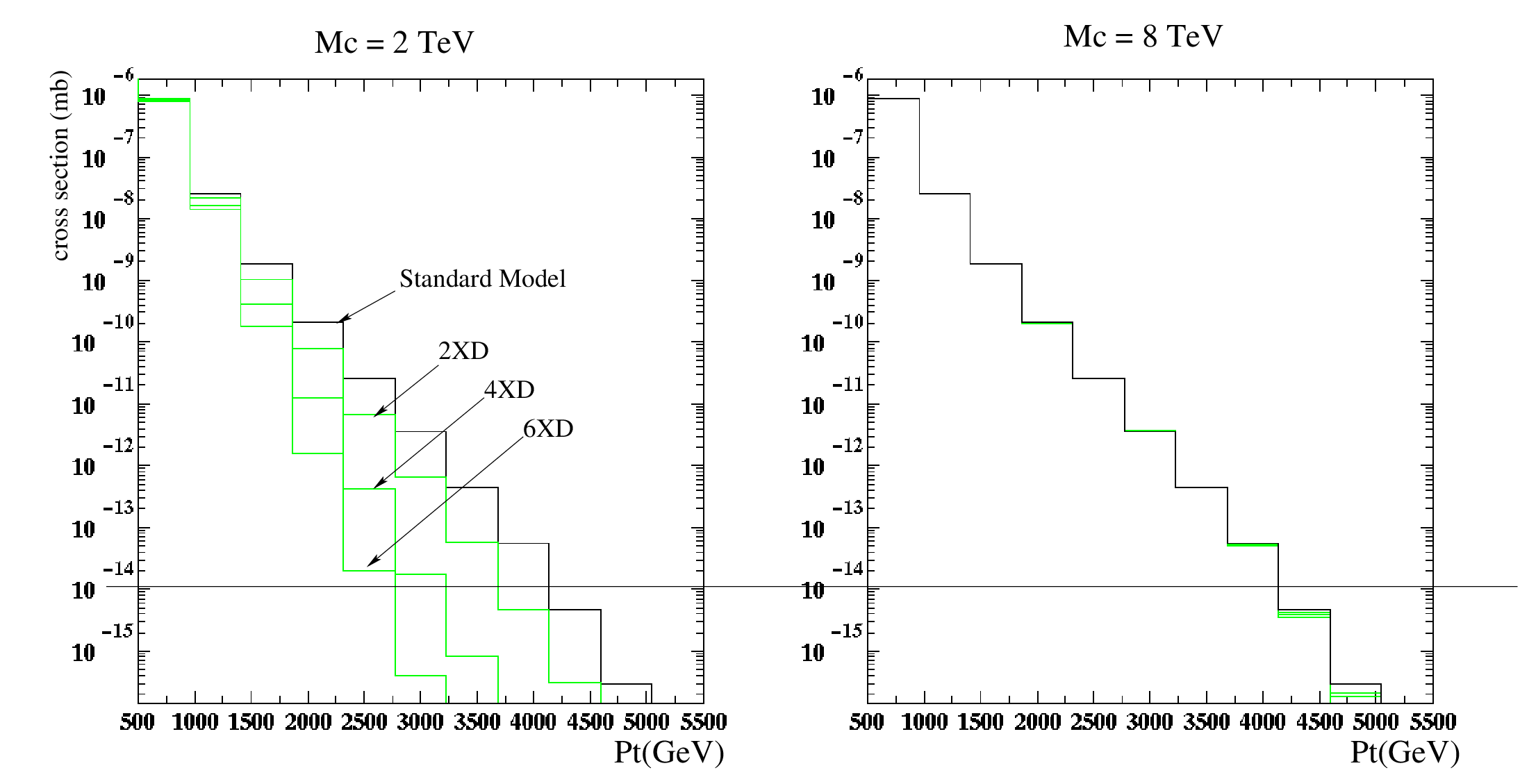,width=0.99\linewidth}
\caption{\it\label{fig:2_8TeV}\djj cross section predictions with compactification scale $M_c$=2 TeV at left
and $M_c$=8 TeV at right. For high $M_c$ values, \xds predictions tend to the Standard Model one. }
\end{center}
\end{figure}

Figures (\ref{fig:2_6xd}) and (\ref{fig:2_8TeV}) compare  the \djs cross section, as a function of Pt, 
for different \xds
parameters to the Standard Model prediction. Figure (\ref{fig:2_6xd}) presents
di-jet cross sections according to the number of \xds ($\delta=2$ and $\delta=6$) and figure (\ref{fig:2_8TeV})
presents them according to the compactification scales ($M_c$=2 and $M_c$=8 TeV).
The Standard Model \djs cross section prediction is presented in 
black and the \xds model prediction in green or red. For a given number of \xds\noindent, predictions for different 
values of the compactification scales 
$M_c$ are presented in figure (\ref{fig:2_6xd}). One can notice a very unusual effect from this 
new physics since
 \xds existence  
reduces the di-jets cross section whereas most other new physics models predict an increase
of the cross section through new particle exchange. Note also that this cancellation effect becomes 
stronger for low compactification scale values. For high compactification scale values,
the theory tends to the Standard Model prediction (figure (\ref{fig:2_8TeV})).
The effect above becomes more important when the number of the \xds increases as shown in figure
(\ref{fig:2_6xd}) for $\delta=6$.\\

\section{\Djs~cross section and proton structure uncertainties}

\subsection{Introduction}

The goal of this section is to estimate the impact of the proton structure uncertainties
on the total \djs cross section. Thus, QCD effects have to be understood in order to
estimate the realistic discovery power of new physics in that channel. \\

 The di-jet cross
section using the factorization theorem can be written as:
\bea
\sigma =\sum_{ij}\int f_i(Q^2,x_i)f_j(Q^2,x_j)\sigma_{i+j\to 2 } \mbox{\rm d }x_i\mbox{\rm d}x_j,
\eea
where $\sigma_{i+j\to 2}$ is the partonic cross section of the di-jet process
$parton_i+parton_j\to parton+parton$. It is calculable at given order in perturbation theory. 
$f_i(Q^2,x_i)$ is the Parton Density
Function which gives the probability of finding a parton $i$ in the proton carrying a fraction
$x_i$ of its momentum. $Q^2$ is the center of mass energy. \\

PDF models with intrinsic errors became available in Jan 2002. The CTEQ collaboration  recently introduced 
a new scheme giving access to evaluate them. The CTEQ6 PDFs \cite{cteq6}.
We implemented this new framework in Pythia 6.152 before the LHAPDF \cite{LHAPDF} interface was public. 

CTEQ6 contains 3 different versions;

\begin{itemize}
\item CTEQ6D: fitted using Deep Inelastic Scatering (DIS) data (BCDMS,
H1,ZEUS,NMC,CCFR,E608 and E866),
\item CTEQ6L: based on Leading Order (LO) calculations and fitted on the the above data
adding the Tevatron ones,
\item CTEQ6M: based on NLO calculations and fitted on the same CTEQ6L data, only this version includes
the uncertainties calculations.  \\ 
\end{itemize}

CTEQ6 fit is based on the Hessian method. This enables a characterization of parton
parametrization in the neighbourhood of the global $\chi^2$ minimum and, then gives access to the
uncertainty calculation. This latter is given as follows\cite{cteq6}:

\begin{itemize}
\item a global fit on the data above is performed using 20 free parameters. This gives 
the nominal PDF set CTEQ6M. The obtained global $\chi^2$ is about 1954 for 1811 data 
points.
\item the global $\chi^2$ is increased by $\Delta\chi^2=100$ to get the error matrix.
\item this matrix is diagonalized to get 20 eigenvectors corresponding to 20 independent
directions in the parameter space.
\item for every eigenvector, up and down excursions are performed in the tolerance gap. This gives
40 sets of new parameters corresponding to 40 new sets of PDFs. These are used in the uncertainty
calculations and are called {\it error PDFs} for simplicity.    \\
\end{itemize}

\Xds sensitivity is calculated by comparing \xd model predictions to the Standard Model one.
As a side product, the Standard Model
prediction zone which contains the Standard Model cross section uncertainty zone is obtained.\\

\subsection{Standard Model Prediction}

\begin{figure}
\begin{center}
\leavevmode
\epsfig{file=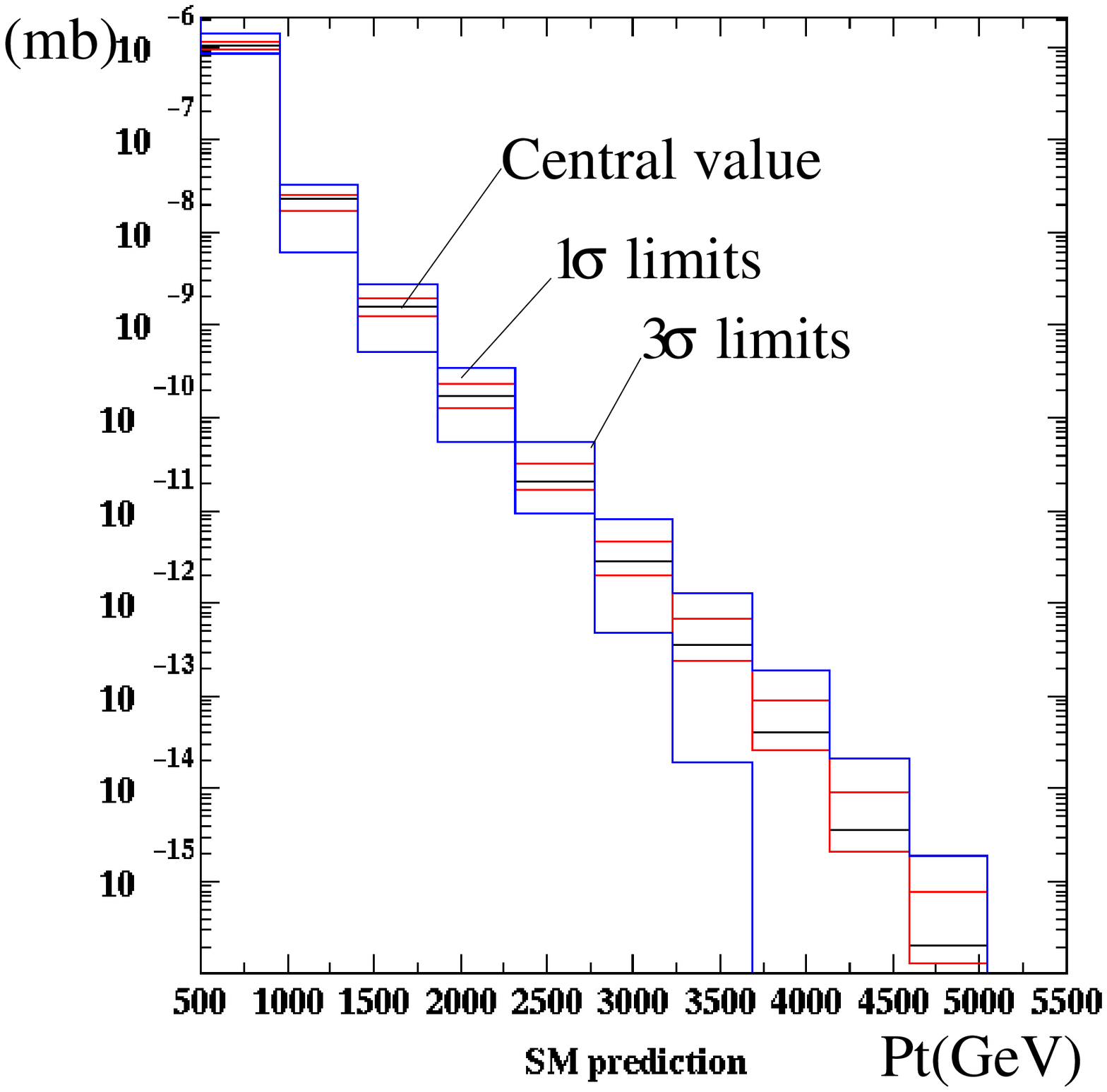,width=0.7\linewidth}
\caption{\it\label{fig:sm_bande}Standard Model \djs cross section prediction and uncertainties zone at
1$\sigma$ and $3\sigma$s.  }
\end{center}
\end{figure}

Let $S_0$ be the PDF set corresponding to the nominal fit, \ie, CTEQ6M. $S_i$
is the set of the error PDFs with $i$ from 1 to 40. $\sigma_0$ is the \djj cross section nominal 
prediction (here, the Standard Model prediction) and $\sigma_i=\sigma(S_i)$ are the cross sections
computed using $S_i$. We use $\Delta \sigma_i^+=\sigma_i-\sigma_0$ when $\sigma_i > \sigma_0$ and 
$\Delta \sigma_i^-=\sigma_i-\sigma_0$ when $\sigma_i < \sigma_0$. The uncertainties are summed quadratically
to define  $\Delta \sigma^\pm=\sqrt{\sum_i \sigma_i^{\pm}}$. The \djs cross section as 
predicted by the Standard Model and its uncertainties is fixed by:
\bea
\sigma^{+ \Delta \sigma^+}_{0 - \Delta \sigma^-}\nonumber
\eea 
and is shown in figure (\ref{fig:sm_bande}) for one and three standard deviations.\\

It is interesting to notice that every measured \djj cross section in this zone is explained within 
the Standard Model by a simple new PDF fit. This interpretation means also that {\it in this zone,
every power of discovering new physics is killed and absorbed by a PDF fit}. One expects a reduction
of the sensitivity to \xds because of these uncertainties.

\subsection{Proton structure uncertainties components}

\begin{figure}
\begin{center}
\leavevmode
\epsfig{file=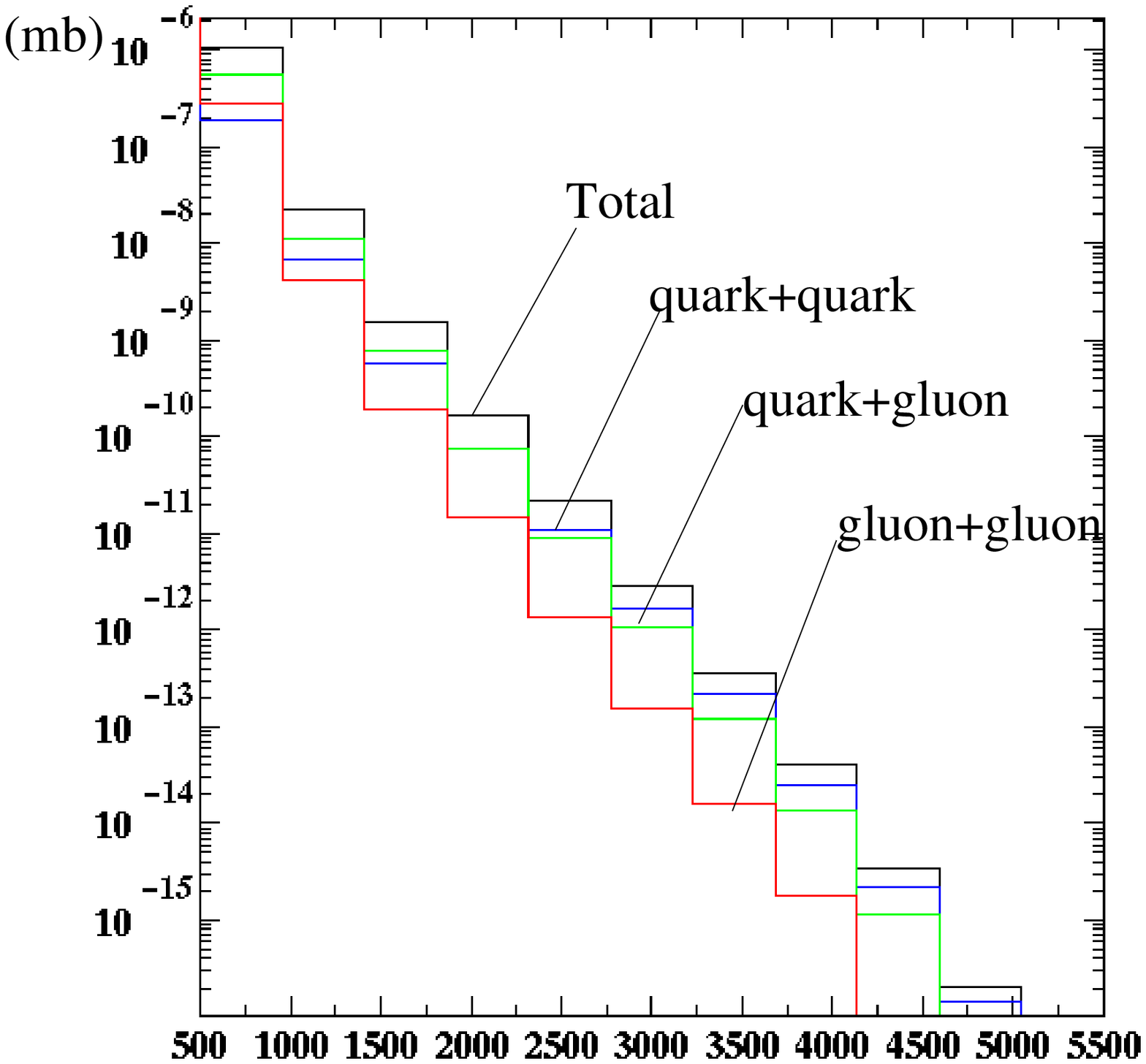,width=0.48\linewidth}
\epsfig{file=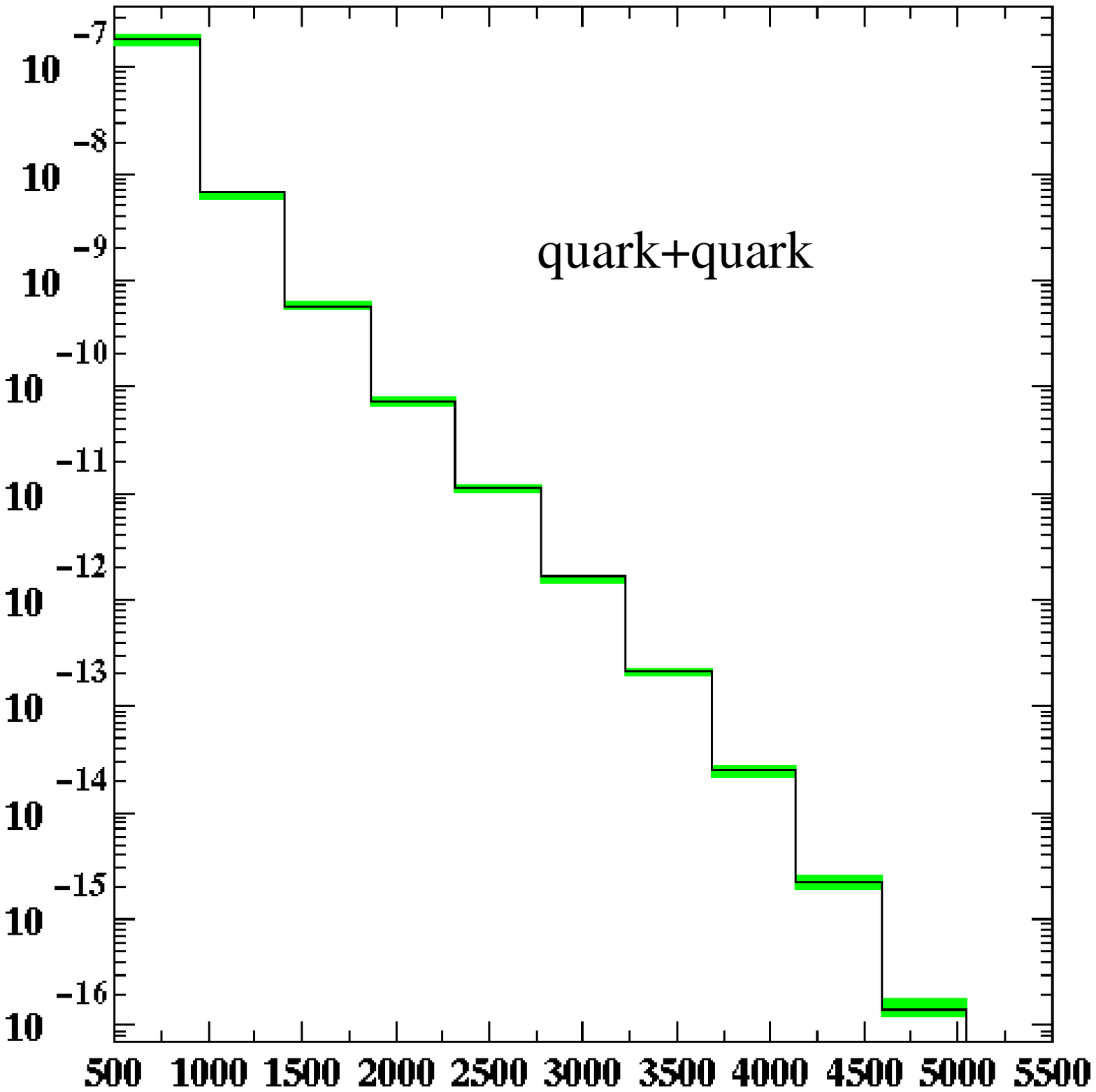,width=0.48\linewidth}
\epsfig{file=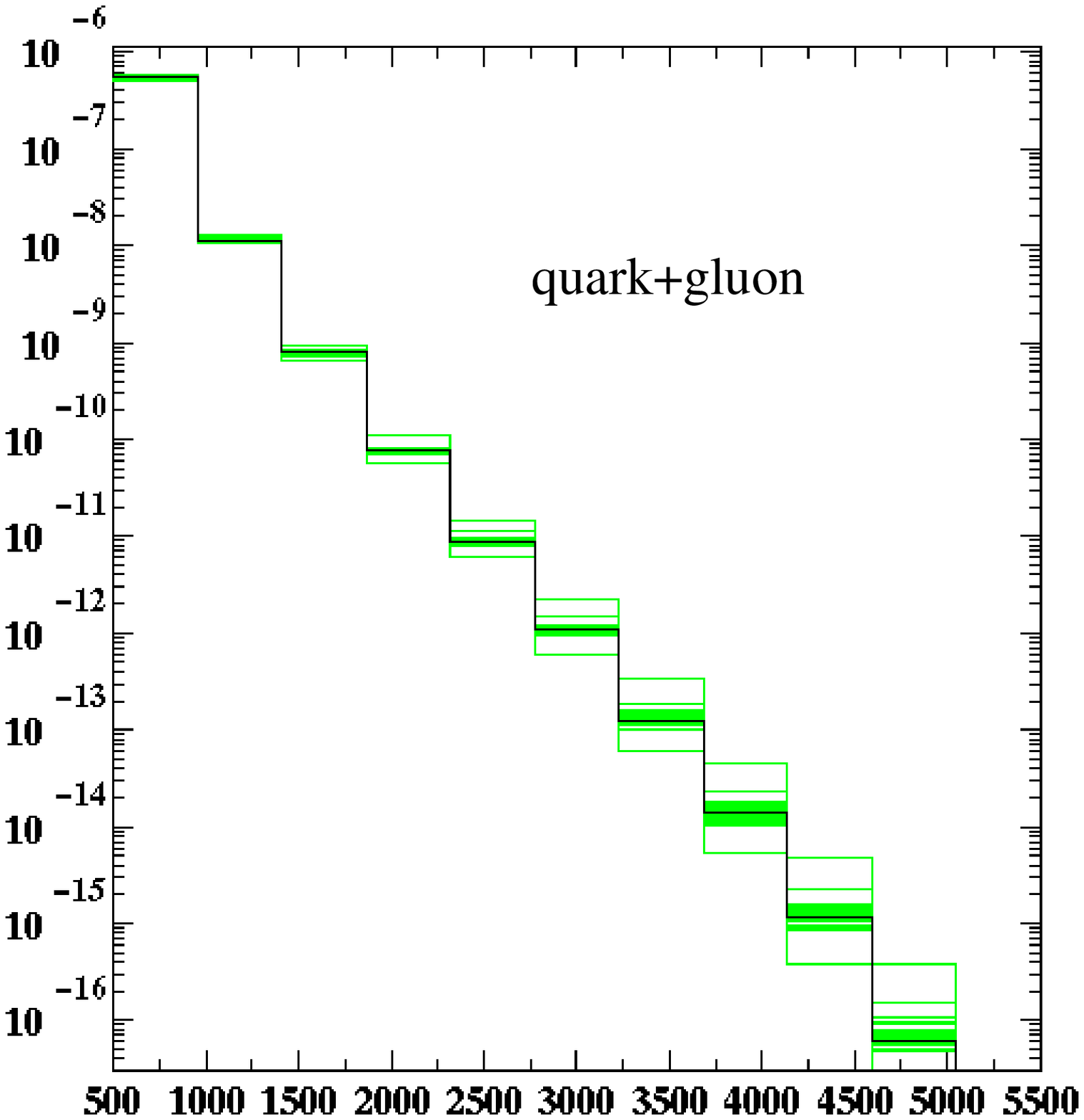,width=0.48\linewidth}
\epsfig{file=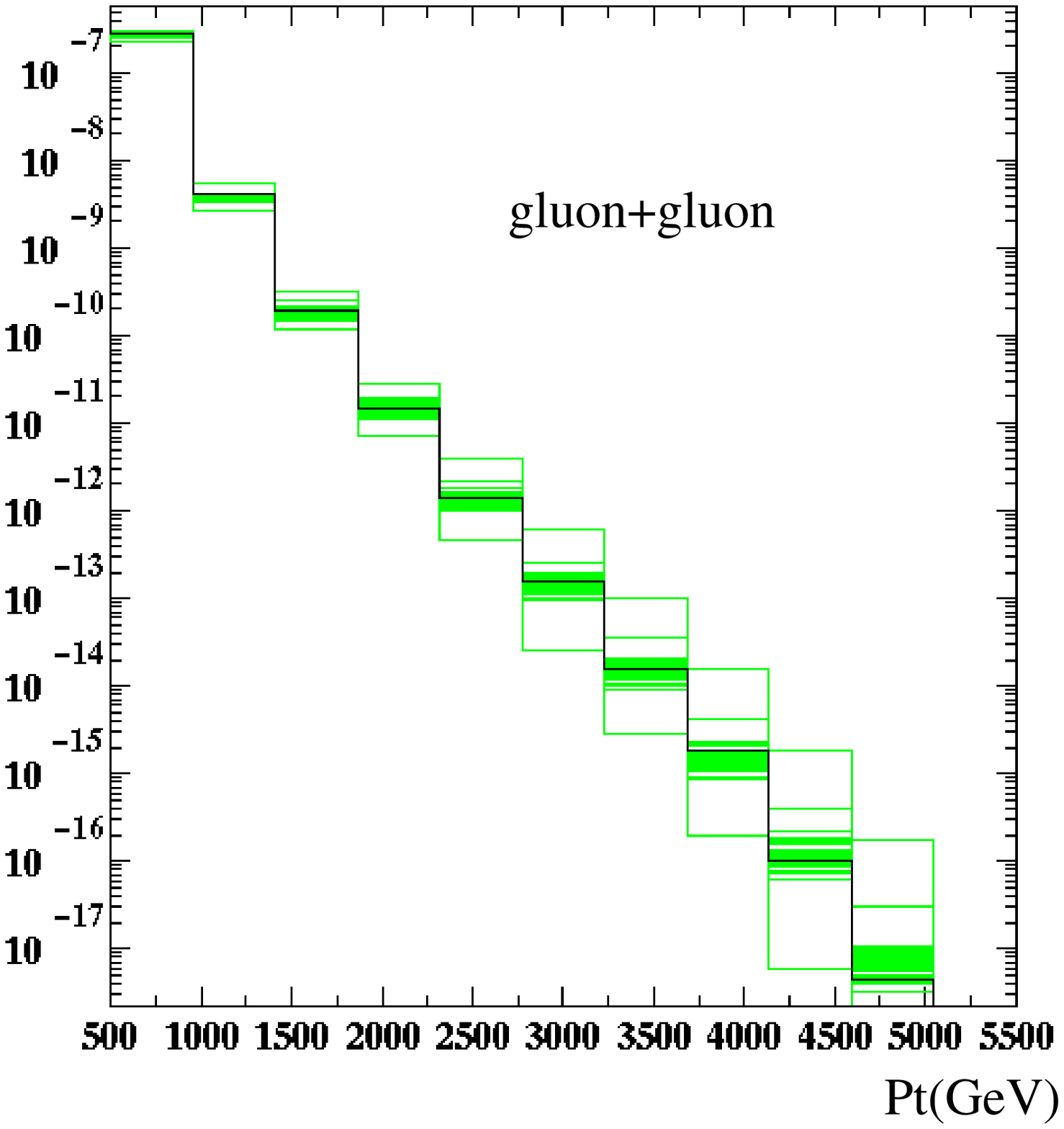,width=0.48\linewidth}
\caption{\it\label{fig:contrib} Standard Model: (up left) different contributions to the total \djj cross 
section according to the nature of the initial partons. (Up right) prediction using CTEQ6M and error PDFs when both
of the initial partons are quarks.(Down left) one of the initial partons is a gluon and the other is a quark.
(Down right) both of the partons are gluons.}
\end{center}
\end{figure}

To understand the source of the large uncertainties above, we show in figure
(\ref{fig:contrib}) the contributions to the $parton+parton\to parton+parton$
cross section depending if the initial partons are quarks, gluons or both.
Figure (\ref{fig:contrib}.b) shows that in the case of the quarks, the 40 PDFs predictions
are confined close to the nominal prediction. This proves that the quark density functions are
well known. This is not the case of the gluon density function. As shown in plots (\ref{fig:contrib}.c)
and (\ref{fig:contrib}.d), when one or both of the initial partons are gluons. It appears that gluons dominates
 the large uncertainties in the predictions.\\ 

\subsection{Impact of the proton structure on the \xds sensitivity}

\begin{figure}
\begin{center}
\leavevmode
\epsfig{file=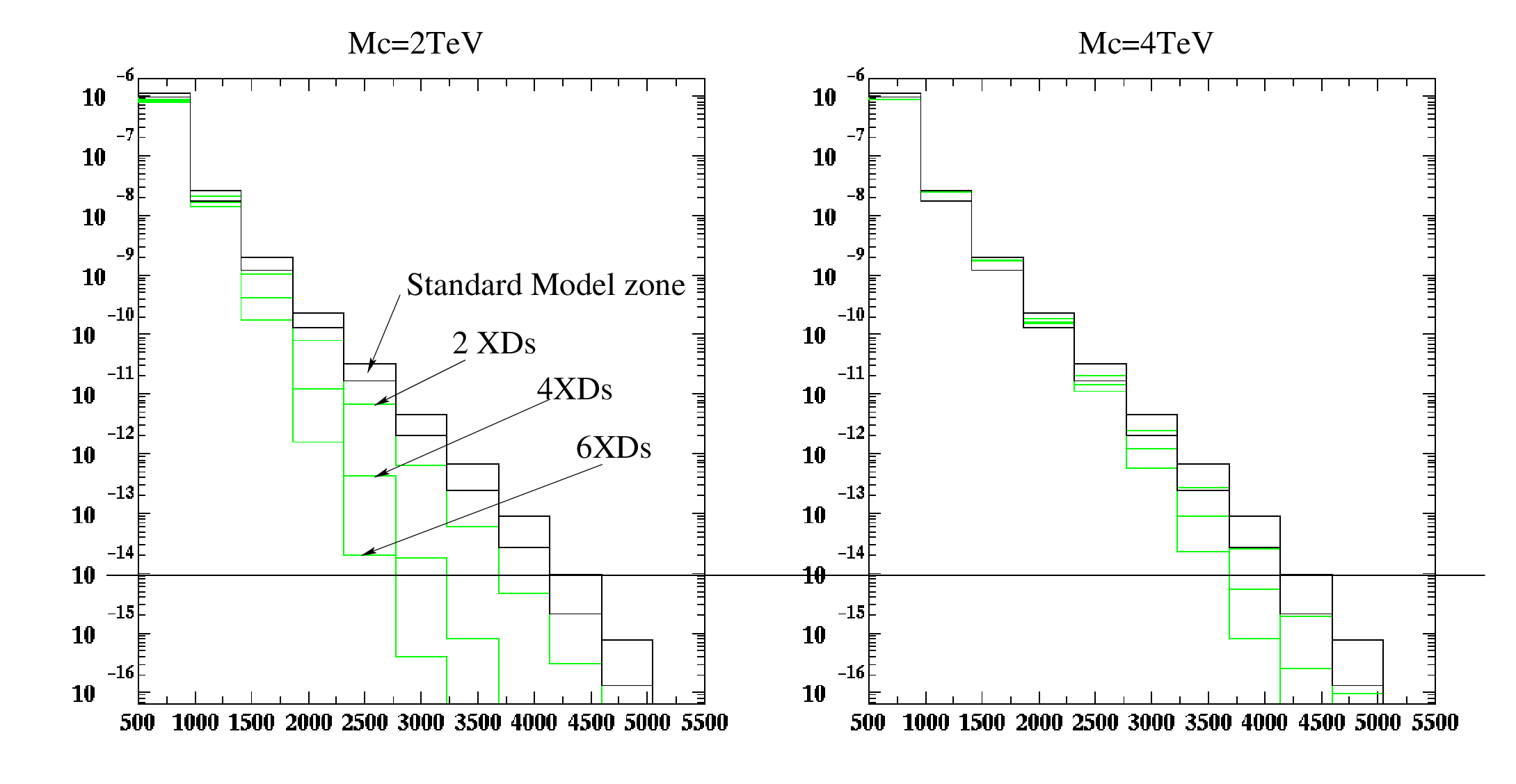,width=0.98\linewidth}
\epsfig{file=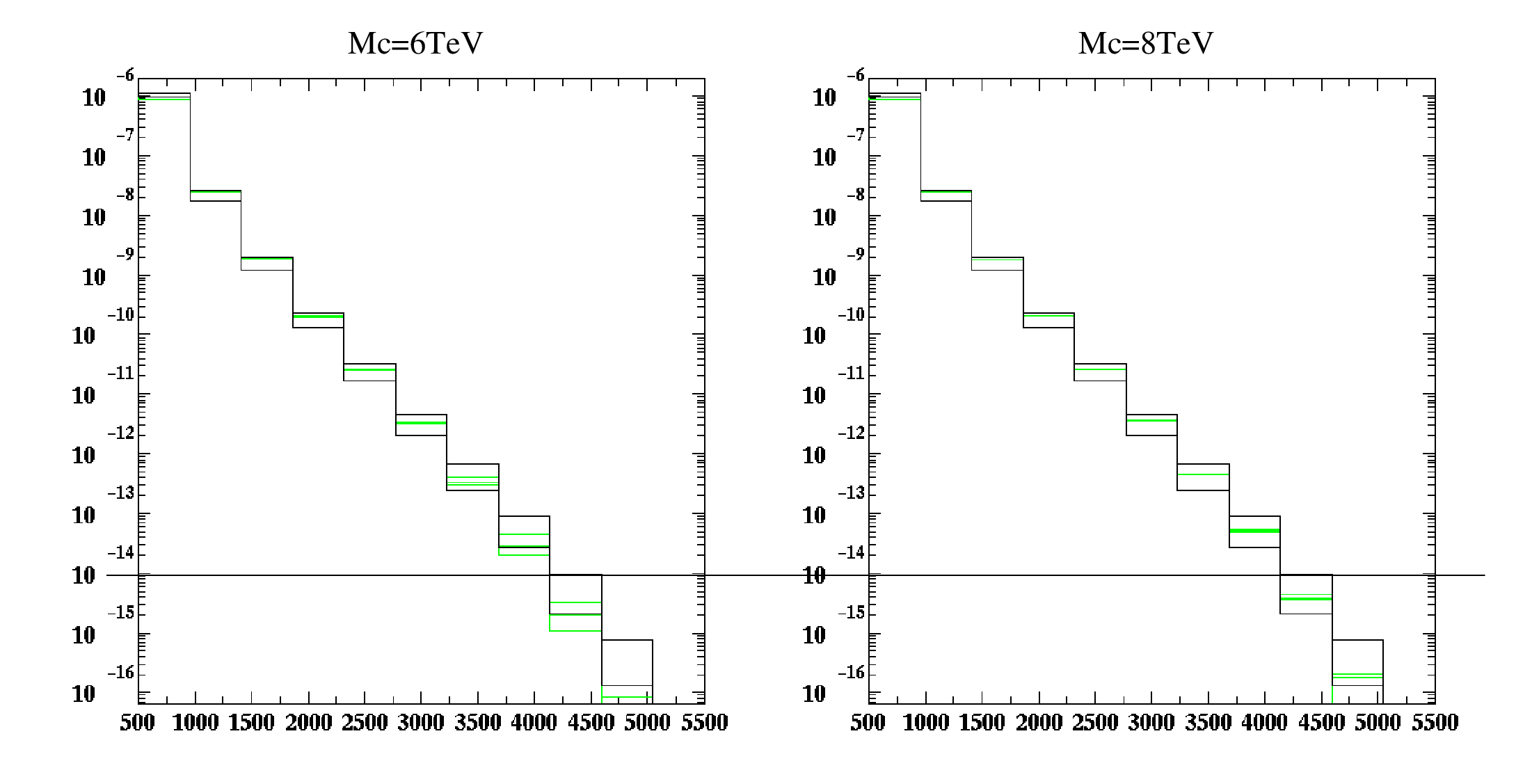,width=0.98\linewidth}
\caption{\it\label{fig:sens_tev} Standard Model and \xds predictions comparison: The \xd predictions are well separated from the Standard
Model uncertainties zone for $M_c=2$ TeV (up left). By increasing the compactification scale, some of those predictions falls into the Standard
Model band and may be considered as Standard Model prediction with a new PDF fit. }
\end{center}
\end{figure}

The Standard Model prediction zone have been established in the previous section and the main PDF source of 
uncertainties is understood. To illustrate the impact of PDF uncertainties on the \xds sensitivity,
\xds predictions are compared to the Standard Model prediction zone.
When we increase the compactification scale $M_c$ the \xds predictions go towards the Standard Model zone and falls into this
latter for $M_c> 2$ TeV, see figure (\ref{fig:sens_tev}). In this case, an \xds prediction is considered
as a Standard Model prediction with a new set of PDFs because it is sitted in the Standard Model prediction zone.
Thus, even if the \xds exist, the corresponding \djs cross section measured in this uncertainties zone is
compatible with the Standard Model prediction and one can not confirm their existence. This fact attenuates the \xds discovery 
potential. The corresponding sensitivity calculations are given below.

\section{Results}

Including the PDF uncertainties, the \xds significance is estimated by comparing the \xds predictions 
to the lower limit of the Standard Model zone because the \xds effect is to cancel the \djs signal.
The statistical significance $S$ is computed for the first year of the LHC high luminosity, 
\ie, 100 fb$^{-1}$. This is given by:
\bea
S=\frac{N^{SM}_{low}-N^{XD}_{(\delta,M_c)}}{\sqrt{N^{SM}_{low}}},
\eea
where $N^{SM}_{low}$ is the lower number of events predicted by the Standard Model. $N^{XD}_{(\delta,M_c)}$
is the number of the events predicted by the \xds model using parameters \xdpar.\\

\begin{figure}
\begin{center}
\leavevmode
\epsfig{file=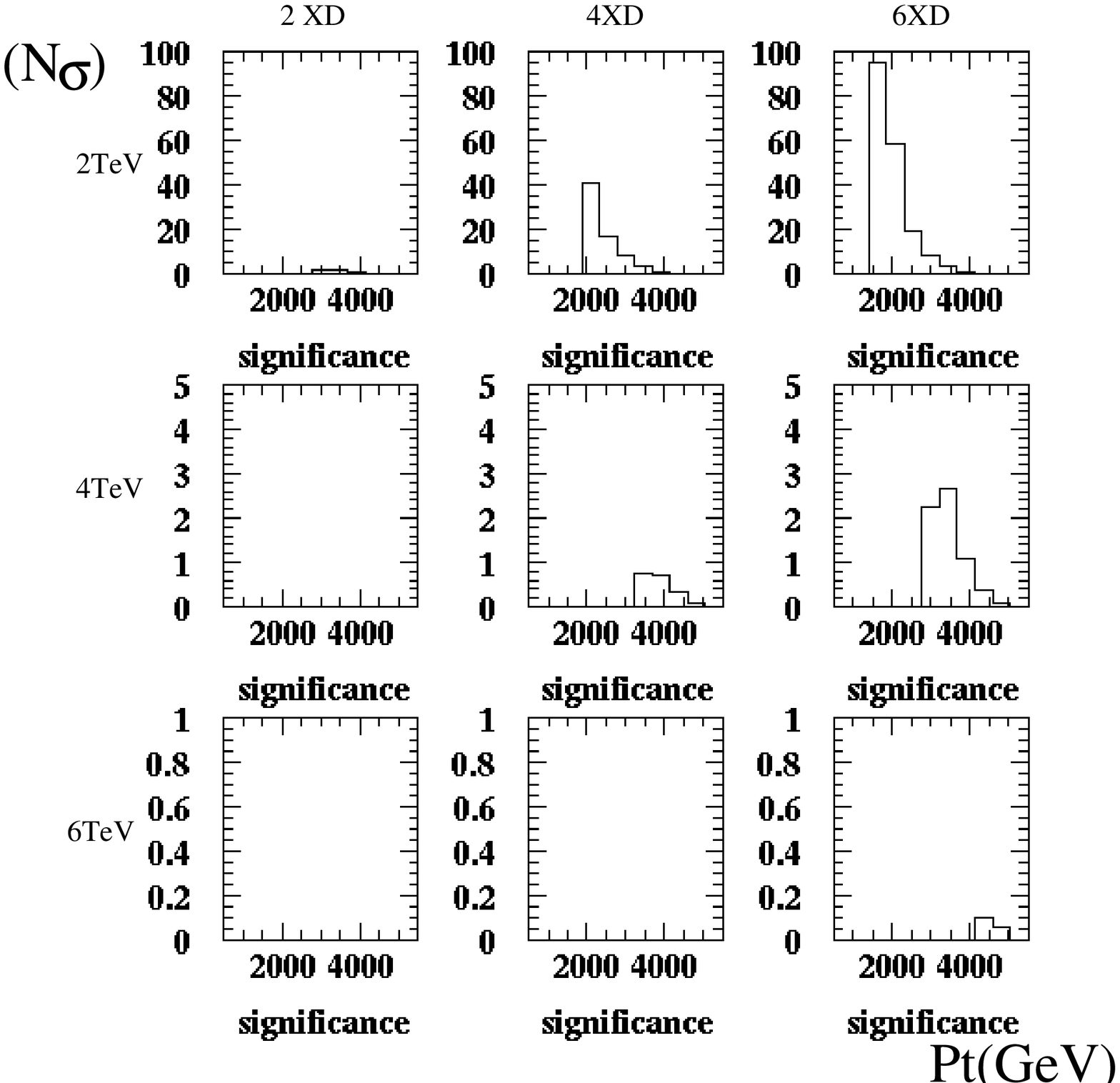,width=0.99\linewidth}
\caption{\it\label{fig:significance} Significance of \xds existance for different numbers of \xds and different
compactification scales. With $\delta=2$ \xds, the significance is lower than $5\sigma$s (left column). 
When we fix the significance sensitivity to more than $5\sigma$, this latter is 
up to 2 TeV in compactification scale for $\delta=4$ \xds (middle column) and between 2 and 4 TeV
in compactification scale for $\delta=6$ \xds (right column). }
\end{center}
\end{figure}
\begin{table}
\caption{\label{tab:significance}\it Upper limit in compactification scale reached by the sensitivity to \xds including 
and without including the proton structure uncertainties. The discovery potential is fixed for a value of the significance $S > 5 \sigma$. }
\begin{center}
\begin{tabular}{|c|c|c|c|}
\hline
&2 & 4 &6 \\
&\xds&  \xds& \xds\\
\hline
\hline
Theoretically&5 TeV&5 TeV&5 TeV\\
 &&&\\
including PDF&$<$ 2 TeV &$<$ 3 TeV&$<$ 4 TeV\\
uncertainties&&&\\
\hline
\end{tabular}
\end{center}
\end{table}
Figure (\ref{fig:significance}) and table (\ref{tab:significance}) show the results of the significance of \xds existance using the
references \cite{Dienes:1998vh,Dienes:1999vg} model. For different \xds \xdpar parameters, we see
that in the case of $\delta=2$, sensitivity is completely lost, ie, the reached upper limit  of $M_c$ is lower than
2 TeV. The discovery potential fixed at $S > 5 \sigma$ reaches $M_c$=2 or 3 TeV for  $\delta=4$ or 6 \xds\noindent. 
Theoretically \cite{balazs}, sensitivity reaches
$M_c$=5 TeV for all the $\delta$ values in the same scenario. So, the ignorance of the proton structure degrades 
significatively 
the discovery of the \xds using the \djj cross section measurement.



\section{Conclusion}

 The discovery of \xds through RGE violation needs a good understanding
of QCD effects at LHC. The analysis presented here gives as a first result the Standard Model \djj cross 
section prediction zone which is an \xds model independent result. 
It shows also as a second result a significant decrease of the \xds discovery potential 
from $M_c=5$ TeV to below 2 to 3 TeV depending on the number of \xds\noindent.
Consequently, using \djs cross section to search for \xds suffers from the  ignorance of the proton structure.
More complicated strategies to recover an important part of the sensitivity have to be developed.
An example of these strategies could be the investigation of the ratio 2 jets/3 jets 
where we can expect a lower sensitivity to the PDF uncertainties. \\
 
High $x$ gluon is responsible of the largest part of the ignorance of the proton structure.
So, it is very important to do QCD analysis
including LHC data to investigate the proton structure and 
to allow LHC experiments to look for new physics. \\

\section*{Acknowledgments}

We thank very much Joey Huston and Daniel Stamp from CTEQ collaboration for many interesting discussions.

\clearpage

\end{document}